\documentclass[]{spie}  %>>> use for US letter paper
 \def\farcs{\hbox{$.\!\!^{\prime\prime}$}}

\newcommand\arcsec{\mbox{$^{\prime\prime}$}}% 
 
\usepackage{amsmath,amsfonts,amssymb}
\usepackage{graphicx}
\usepackage[colorlinks=true, allcolors=blue]{hyperref}
\usepackage{caption}
\usepackage{subcaption}

\title{A new type of exoplanet direct imaging search: the SCExAO/CHARIS survey of accelerating stars\thanks{~~Based on data collected at Subaru Telescope, which is operated by the National Astronomical Observatory of Japan.}}

\author[a,b,c]{Thayne Currie}
\author[d]{Timothy D. Brandt}
\author[e]{Masayuki Kuzuhara}
\author[f]{Jeffrey Chilcote}
\author[g,c]{Edward Cashman}
\author[h,c]{R. Y. Liu}
\author[i]{Kellen Lawson}
\author[f]{Taylor Tobin}
\author[d]{G. Mirek Brandt}
\author[b,e,j,k]{Olivier Guyon}
\author[b]{Julien Lozi}
\author[b]{Vincent Deo}
\author[b]{Sebastien Vievard}
\author[b]{Kyohoon Ahn}
\author[b,l,m]{Nour Skaf}
\affil[a]{NASA-Ames Research Center, Moffett Field, California, USA}
\affil[b]{Subaru Telescope, 650 N. Aohoku Pl., Hilo, Hawai'i, USA}
\affil[c]{Eureka Scientific, Oakland, California, USA}
\affil[d]{University of California-Santa Barbara, Santa Barbara, CA, USA}
\affil[e]{Astrobiology Center of NINS, 2-21-1, Osawa, Mitaka, Tokyo, Japan}
\affil[f]{University of Notre Dame, South Bend, IN, USA}
\affil[g]{University of Hawai`i-Hilo, Hilo, HI, USA}
\affil[h]{Columbia University, New York, NY, USA}
\affil[i]{University of Oklahoma, Norman, Oklahoma, USA}
\affil[j]{Steward Observatory, University of Arizona, Tucson, AZ, USA}
\affil[k]{James C. Wyant College of Optical Sciences, University of Arizona, Tucson, AZ, USA}
\affil[]{LESIA, Observatoire de Paris, Université PSL, CNRS, Sorbonne Universit\'e, Universit\'e de Paris, 5 place Jules Janssen, 92195 Meudon, France}
\affil[m]{Department of Physics and Astronomy, University College London, London, United Kingdom}
\authorinfo{Further author information: (Send correspondence to T. Currie)\\T. Currie: E-mail: currie@naoj.org}

\begin{document} 
\maketitle

\begin{abstract}
We present first results from a new exoplanet direct imaging survey being carried out with the \textit{Subaru Coronagraphic Extreme Adaptive Optics} project (SCExAO) coupled to the CHARIS integral field spectrograph and assisted with Keck/NIRC2, targeting stars showing evidence for an astrometric acceleration from the \textit{Hipparcos} and \textit{Gaia} satellites.   Near-infrared spectra from CHARIS and thermal infrared photometry from NIRC2 constrain newly-discovered companion spectral types, temperatures, and gravities.   Relative astrometry of companions from SCExAO/CHARIS and NIRC2 and absolute astrometry of the star from \textit{Hipparcos} and \textit{Gaia} together yield direct dynamical mass constraints.   Even in its infancy, our survey has already yielded multiple discoveries, including at least one likely jovian planet.    We describe how our nascent survey is yielding a far higher detection rate than blind surveys from GPI and SPHERE, mass precisions reached for known companions, and the path forward for imaging and characterizing planets at lower masses and smaller orbital separations than previously possible. 
\end{abstract}


\section{Introduction}
\indent Over the past decade, facility and \textit{extreme} adaptive optics (AO) systems on 8-10m class ground based-telescopes have provided the first direct detections of superjovian extrasolar planets \cite{Marois2008,Marois2010,Lagrange2010,Currie2014,Currie2015,Macintosh2015,Keppler2018,Bohn2021}.  The vast majority of imaged exoplanets orbit at wide separations from their host star, typically in systems with ages between 1 and 100 Myr.   Follow-up near to mid infrared (IR) photometry and spectroscopy have revealed key insights into the clouds, chemistries, and gravities of young gas giants \cite{Currie2011,Barman2015,Rajan2017}.    Dedicated exoplanet direct imaging surveys have shed light on the frequencies of jovian exoplanets beyond 10 au, as a function of stellar mass, and around stars surrounded by cold, Kuiper belt-like debris disks \cite{Galicher2016,Meshkat2017,Nielsen2019,Vigan2020}.  

Unfortunately,  the low yield of \underline{\textit{blind}} direct imaging surveys show that exoplanets imageable with current instruments are rare beyond 10 au \cite{Nielsen2019}.   For example, out of $\sim$ 300 young systems surveyed, the GPIES survey has only 9 imaged substellar companions around $\sim$ 6 stars: exoplanets around three of these systems (HR 8799, $\beta$ Pic, HD 95086) were already known prior to the survey. Companions detected by these current blind imaging surveys are typically more than 2--5 $M_{\rm J}$ in mass and orbit well beyond $\sim$ 3 au, where the jovian planet frequency peaks \cite{Fernandes2019,Fulton2021,Rosenthal2021}.  Absent substantial contrast gains enabling the detection of mature reflected-light planets -- e.g. 10$^{-8}$ at 0\farcs{}1--0\farcs{}5 in the near IR -- future blind surveys will also have a low yield.   Improving the yield of blind surveys will only be possible with substantial contrast gains at small separations and/or sensitivity gains at wider separations \cite{Crepp2011}.
 % (e.g. Crepp \& Johnson 2011).   
These low yields and corresponding sparse coverage of ages, temperatures, and surface gravities for imaging discoveries impede our understanding of the atmospheric evolution of gas giant planets. 

In comparison to blind searches, \underline{targeted} searches focusing on stars showing evidence for the gravitational pull from a massive jovian planet will likely yield far more detections.   Recent targeted, radial-velocity selected surveys conducted with facility (conventional) AO -- e.g. the TRENDS survey -- offer a proof-of-concept\cite{Crepp2016}, yielding discoveries of multiple very low mass stars and brown dwarfs.   Astrometric monitoring can identify young stars with likely imageable planets, especially those whose activity and spectral types preclude precise radial-velocity measurements.    

The {\it Hipparcos}-{\it Gaia} Catalog of Accelerations (HGCA), recently updated to include \textit{Gaia}-eDR3 data, provides absolute astrometry for 115,000 nearby stars, including hundreds of nearby systems showing clear dynamical evidence for unseen substellar companions\cite{Brandt2021}.   A direct imaging survey targeting stars in HGCA using an \textit{extreme} AO system should result in a far higher yield of newly discovered brown dwarfs and extrasolar planets. Accelerations derived from the HGCA can provide dynamical masses of known imaged exoplanets and low-mass brown dwarfs independent of luminosity evolution models and irrespective of uncertainties in stellar ages\cite{Brandt2019}.  

In this paper, we describe the SCExAO/CHARIS direct imaging survey of accelerating stars drawn from the HGCA.    Section \S{2} summarizes the current state of SCExAO/CHARIS and soon-to-be implemented upgrades allowing it to image planets lower in mass ($\lesssim$2 $M_{\rm J}$) and smaller in semimajor axis ($a$ $\sim$ 3--5 au) than possible with any ground-based system.   We describe the data reduction pipeline used for this survey -- the now-public CHARIS data processing pipeline (DPP) -- and efforts to improve its speed, increase its versality, and widen its user base with an upcoming translation from IDL to Python.   We then display SCExAO/CHARIS's sensitivity to self-luminous superjovian planets and we discuss how survey observations with Keck/NIRC2 NIRC2 in the thermal IR using the observatory's new near-IR Pyramid wavefront sensor complement SCExAO/CHARIS data, especially for intermediate-aged systems (Section \S{3}). In Section \S{4}, we discuss target selection for our survey, focusing on stars whose youth, distances, and accelerations are indicative of imageable substellar  companions.   We summarize discoveries made during the first full year of this survey in Section \S{5} and illustrate the full characterization power of the survey in Section \S{6}.   Finally, we forecast the future of this program with the soon-to-be upgraded SCExAO, future \textit{Gaia} data releases, and astrometric planet detection from the \textit{Roman Space Telescope} (\S{7}).

\begin{figure}[ht]
   \begin{center}
  % \begin{tabular}{c} %% tabular useful for creating an array of images 
  \centering
   \includegraphics[scale=0.8,clip]{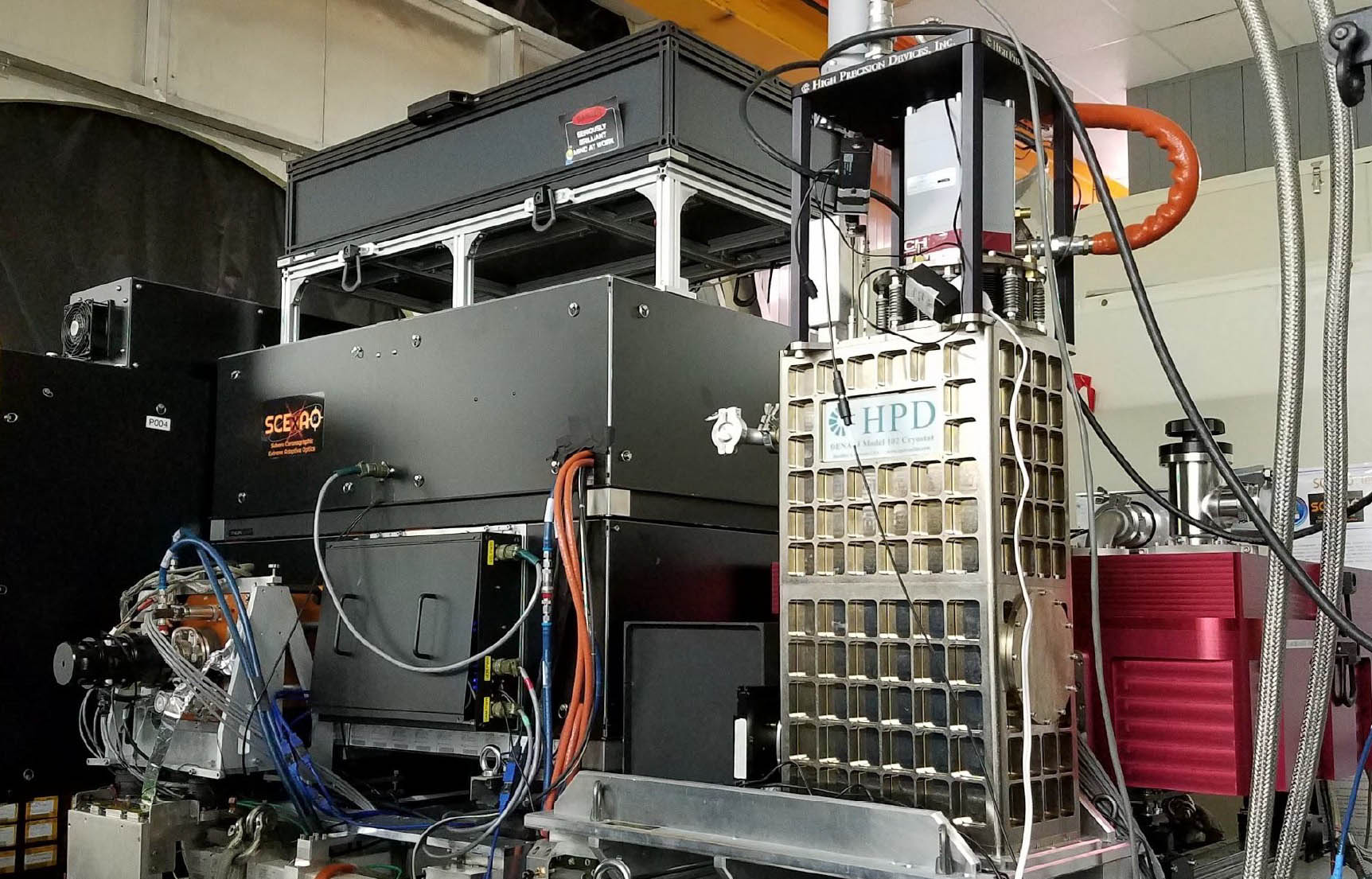}
  % \end{tabular}
   \end{center}
   \caption
%>>>> use \label inside caption to get Fig. number with \ref{}
   { \label{fig:schematic} 
       SCExAO (black box, left) and CHARIS (red instrument) on the Nasmyth platform at Subaru.   The tall grey tower is the MKIDS Exoplanet Camera (MEC), which often operates simultaneously with CHARIS, taking light in $Y$ band (1.03 $\mu m$).   
       %The current schematic of SCExAO.  Note that the coronagraphic low-order loop is not in normal operation; the MKIDs camera (MEC) is undergoing commissioning.
}
\end{figure}
\section{The SCExAO/CHARIS High-Contrast Imaging Platform}
Currie et al. \citenum{Currie2020c} details the hardware and software used for wavefront control with SCExAO and integral field spectroscopy with CHARIS (shown in Figure \ref{fig:schematic}).   Here, we summarize key aspects of SCExAO, CHARIS, and reduction of SCExAO/CHARIS data relevant to our survey.   

\subsection{SCExAO} 
SCExAO \footnote{\url{http://www.naoj.org/Projects/SCEXAO/index.html}} is a cutting-edge extreme adaptive optics platform designed to image and characterize the spectra of young jovian planets at solar system-like scales (Jovanovic et al. \citenum{Jovanovic2015}; Lozi et al. \citenum{Lozi2018SCExAO}; K. Ahn 2021, SPIE proceedings).  As described in Currie et al. \citenum{Currie2020c}, SCExAO utilizes a modulated Pyramid wavefront sensor (PyWFS) sending light to an EMCCD camera (OCAM$^{2}$K from First-Light Imaging) operating in the red optical (typically 600-900 nm) to sense the incoming aberrated wavefront.   A 2000-actuator Boston MicroMachines MEMS deformable mirror (DM) corrects for these aberrations.   While the SCExAO wavefront control loop can run at speeds of 3.5 kHz, we typically operate it at a slower 2 kHz, combining the main loop with predictive wavefront control\cite{MalesGuyon2018} to substantially reduce temporal bandwidth errors that limit contrasts at small angular separations and wavefront sensor noise limiting contrast at mid spatial frequencies.   The SCExAO control loop uses a mix of CPU and GPU resources, configured to a Real-Time Controller computer system to achieve extremely low-latency wavefront control.

As SCExAO's DM stroke is too small to fully correct for the wavefront error induced by atmospheric turbulence,  Subaru's facility AO system (AO-188) provides an initial woofer correction for SCExAO, typically yielding $\sim$ 30--40\% Strehl in H band.  In early 2022, we will replace AO188's DM with a 3200-actuator magnetic DM (ALPAO DM3228) and upgrade its wavefront sensor from a 188-element avalanche photodiode array to a fast low noise imaging array.   We will also implement a near-infrared PyWFS option \cite{2019PASP..131d4503L} similar to that recently installed at Keck \cite{Bond2020}.
%We plan to replace Subaru's venerable facility AO system, which uses a curvature wavefront sensor driving a DM with only 188 actuators.   AO-188's low actuator density across the telescope pupil allows a modest correction of atmospheric turbulence and exhibits strong vibration modes precluding full-speed operation\cite{Oya2006}.   Its successor will operate with a much faster (2 kHz) and far higher-order DM (3200 actuators).   Our plan is to upgrade its wavefront sensor from the current 188-element avalanche photodiode array to an EMCCD.   The camera upgrade will allow the ``AO-3000" to substantially reduce both the fitting error and temporal bandwidth error, yielding $H$ band Strehl ratios approaching 85\% and a reduction in the PSF halo by nearly a factor of 5.    

\begin{figure}[ht]
   \begin{center}
  % \begin{tabular}{c} %% tabular useful for creating an array of images 
  \centering
   \includegraphics[width=0.925\textwidth,clip]{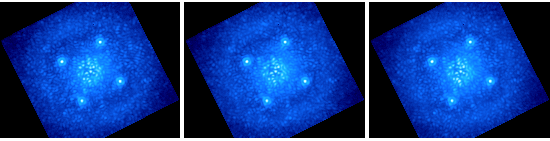}
  % \end{tabular}
   \end{center}
   \vspace{-0.1in}
   \caption
%>>>> use \label inside caption to get Fig. number with \ref{}
   { \label{fig:darkhole} 
   SCExAO/CHARIS wavelength slices ($\lambda_{\rm o}$ = 1.16 $\mu m$) obtained from three consecutive exposures, showing diffraction-limited satellite spots and a dark hole within 0\farcs{}7 ($\sim$23 $\lambda$/D).
   %Laboratory-based response matrix calculations for SCExAO blend a) Hadamard pokes sampling high spatial frequencies and b) Zernike/Fourier pokes sampling low spatial frequencies.   This approach, when combined with predictive control and in lieu of acquiring an RM on sky, yields far superior control of low-order modes and diffraction-limited imaging in the red optical (panel c).   Note that the rotation angle for the DM and PyWFS are offset by 45$^{o}$, explaining the tilt of the poke vs. PyWFS response patterns.
       %The current schematic of SCExAO.  Note that the coronagraphic low-order loop is not in normal operation; the MKIDs camera (MEC) is undergoing commissioning.
}
\end{figure}

SCExAO achieves Strehl ratios (SR) in $H$ band (1.65 $\mu m$) of $\sim$ 80-90\% in median conditions, up to SR $\sim$ 93--94\% for the brightest stars observed in the best conditions \cite{Currie2019b} (Figure \ref{fig:darkhole}).  Performance for optically fainter stars (I $\sim$ 9--11) is highly dependent on the atmospheric coherence time and, in turn, the performance of AO188.   In poor seeing conditions (e.g. $\theta_{\rm V}$ $\gtrsim$ 1--1.5\arcsec{}), AO188 typically fails to provide a correction stable enough for SCExAO to then achieve bona fide extreme AO performance.  The upcoming DM upgrade will substantially improve achievable raw contrasts for bright stars, allowing extreme AO capability in poor conditions and also substantially improve performance in good/median conditions.   The near-IR PyWFS sensor will allow consistent extreme AO corrections for optically faint, near-IR bright stars (e.g. K and M stars).

Thus, most of our survey currently focuses on optically-bright stars (I $\lesssim$ 8), which typicaly have spectral types, temperatures, and masses comparable to or greater than the Sun.   When conditions are poor, we typically take short snapshot observations of many stars to cull our target list of binaries.  When conditions are median to excellent, we obtain deep sequences of stars without binary companions whose accelerations can only be consistent with substellar companions.

%, a form of dome seeing caused by temperature gradients on opposite sides of the secondary spider when the wind speed is extremely small.   

\subsection{The CHARIS Integral Field Spectrograph}

CHARIS, a lenslet-based cryogenic integral field spectrograph operating in the near-IR $\mu m$\cite{Groff2016}, receives starlight sharpened by SCExAO and suppressed by a suite of coronagraphs.      Spatial dimensions for CHARIS data cubes cover 201 pixels by 201 pixels.  The effective field of view for CHARIS in each channel is a square region rotated by 45$^{o}$ with full 360$^{o}$ coverage out to 1\farcs{}07 (1\farcs{}48 in the corners).     %allows for more detailed, follow-up characterization.
In low-resolution (``broadband") mode, CHARIS integral field spectra consist of 22 channels whose central wavelengths range from 1.16 to 2.37 $\mu m$ at a resolution of $\mathcal{R}$ $\sim$ 20.   CHARIS also offers a higher resolution mode covering J (15 channels), H (20 channels), or K (17 channels) passbands ($\mathcal{R}$ $\sim$ 70).  

The CHARIS detector characteristics and wavelength-dependent image quality determine the standard operational mode for our HGCA survey: observing in low-resolution mode.   Young L and L/T transition exoplanets tend to be redder and fainter at the shortest near-IR wavelengths than old field brown dwarf counterparts due to their thicker clouds \cite{Currie2011}, trends which favor observing at longer wavelengths.   For a given wavefront error, the Strehl ratio is also higher in K band than in J or H: e.g. 125 nm of residual wavefront error corresponds to SR(J,H,K) $\sim$ 0.70, 0.81, and 0.88, respectively.   However, the sensitivity for CHARIS $K$ band spectra is poorer than in low resolution over the wavelengths covering K band ($m_{\rm ZP, 2.2 \mu m}$ = 17.5 in low resolution vs. 15.5 in K band).   T dwarf planets like 51 Eri b have neutral to blue $H$-$K$ colors \cite{Macintosh2015}.   Finally, CHARIS's far broader bandpass in low resolution mode makes speckle supression with \textit{spectral differential imaging} (SDI) far more powerful, as it can be used down to very small angular separations ($\rho$ $\sim$ 0\farcs{}2) with minimal self-subtraction \cite{Currie2018a}.  
%in while the CHARIS covers a 15-20\% bandpass at higher resolution, at lower resolution CHARIS has a 200\% bandpass.  
%After receiving light that is well corrected from SCExAO and partially suppressed by a coronagraph, a sparse image is formed on the lenslet array.   After a pinhole array mitigates lenslet diffraction, light from the lenslet array is dispersed from one of two prisms onto a 2048x2048-pixel Hawaii 2RG detector into 135x135 30 pixel-long microspectra.   With the low-resolution prism in position, CHARIS spectra have a resolution of $\mathcal{R}$ $\sim$ 20 and cover 22 channels with central wavelengths of range 1.15--2.37 $\mu m$.   In its high-resolution mode ($\mathcal{R}$ $\sim$ 70), CHARIS spectra cover the J, H, or K passbands.   The CHARIS Data Reduction Pipeline (DRP) converts raw CHARIS detector data into data cubes ($x$ by $y$ by $\lambda$). %allows for more detailed, follow-up characterization.

 A typical survey night generates 50-100 GB of raw data.  The CHARIS Data Reduction Pipeline (DRP)\cite{Brandt2017} converts raw CHARIS detector data into data cubes ($x$ by $y$ by $\lambda$).   The raw data are typically pulled directly from the CHARIS computer at the end of the night by the principal investigator (PI) and turned into data cubes by using the DRP or from the \textit{ADEPTS} infrastructure hosted at Notre Dame running the DRP on cluster computer\cite{Tobin2020}.  While turnaround time for PI-based extractions vary, \textit{ADEPTS} typically produces a full night's set of cubes ready for post-processing within hours. Although processing by \textit{ADEPTS} is currently performed at request, when fully operational, CHARIS data will be processed automatically.

\subsection{The CHARIS Data Processing Pipeline}
The CHARIS Data Processing Pipeline (DPP) has been used in nearly all SCExAO/CHARIS papers thus far \cite{Currie2018a,Currie2018b,Goebel2018,Rich2019,AsensioTorres2019,Currie2019,Uyama2020a,Lawson2020,Currie2020a} and is utilized in our HGCA survey.  The DPP performs sky subtraction, cube registration, spectrophotometric calibration, PSF subtraction, and planet and disk forward-modeling.  From sequence-combined data, it then yields contrast curves, extracted spectra of companions, and analyses comparing the spectra to other low mass objects and to atmosphere models.

T.C's Github page hosts a public release version of the DPP written in IDL for the wider SCExAO/CHARIS user base: \url{https://github.com/thaynecurrie/charis-dpp}.  In addition to processing CHARIS total intensity data, the DPP now processes integral field polarimetry mode data (discussed in K. Lawson et al. 2021, SPIE).    Full Github-based documentation and tutorials will be added by late 2021.   A separate repository hosts sample CHARIS data: the 31 August HD 33632 data responsible for confirming the brown dwarf HD 33632 Ab \cite{Currie2020a} and will include other data sets later this year: \url{https://github.com/thaynecurrie/charis_data_sets}.
The Github page and Currie et al. \citenum{Currie2020c} give a brief walkthrough of key DPP reduction steps.  In most cases, calling key modules without arguments yields usage instructions similar to docstrings in Python.  

From the delivery of rectified data cubes, a typical CHARIS sequence can be fully processed by the DPP in under an hour clock time with quick-look results (i.e. the identification of a planet candidate) in 15-30 minutes.  Thus, candidate companions identified identified from a previous night's data can be full identified and prioritized for a following night.

To broaden the base of survey team members able to reduce SCExAO/CHARIS data and increase the speed and versatility of the pipeline, we are now converting it to Python 3 (lead by R.L., E.C., K.L. and T.C).  While the IDL-based code relies significantly on the IDL astronomer's users library, the Python 3 version will utilize a combination of custom modules/functions and widely used packages (\texttt{astropy}, \texttt{scipy}, \texttt{numpy}, \texttt{numba}, etc.).   The full Python translation of the DPP will enable much faster multi-threading and parallelized implementation of PSF subtraction methods (e.g. ADI/SDI+A-LOCI) and new advances that exploit forward-modeling to achieve deeper contrast\cite{Ruffio2017}.  The Pythonized DPP can also be used as a part of a real-time reduction of CHARIS data\cite{Tobin2020}.

\begin{figure}[h]
\vspace{-0.1in}
   \begin{center}
  % \begin{tabular}{c} %% tabular useful for creating an array of images 
  \centering
   \includegraphics[scale=0.47,clip]{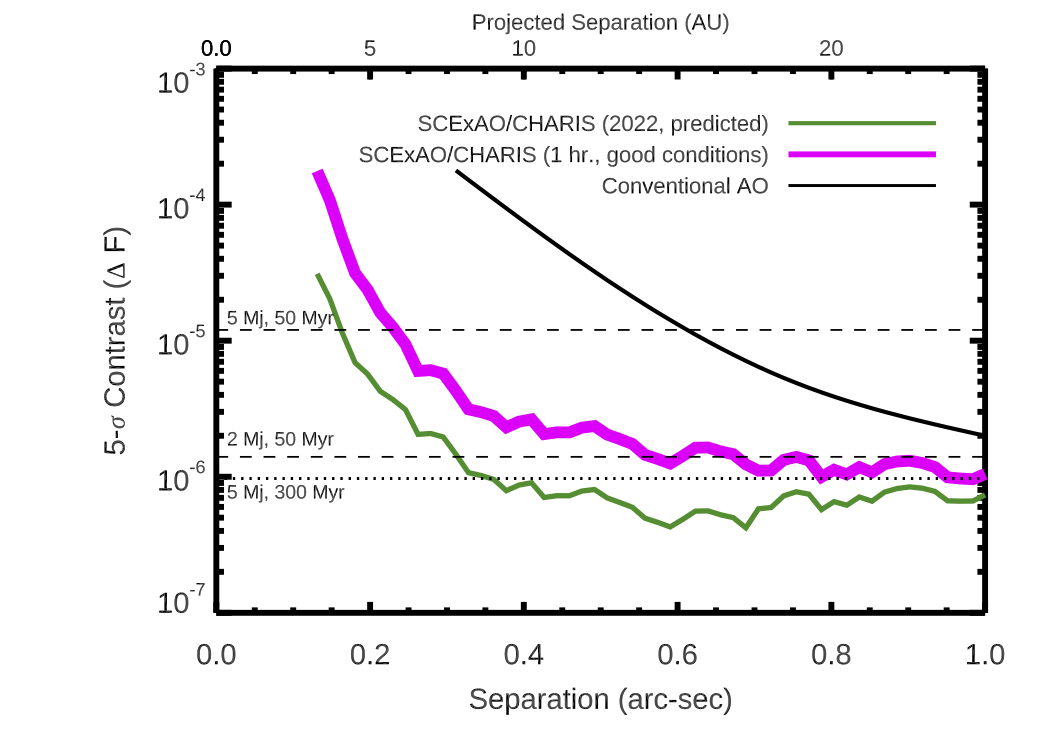}
     \includegraphics[scale=0.47,clip]{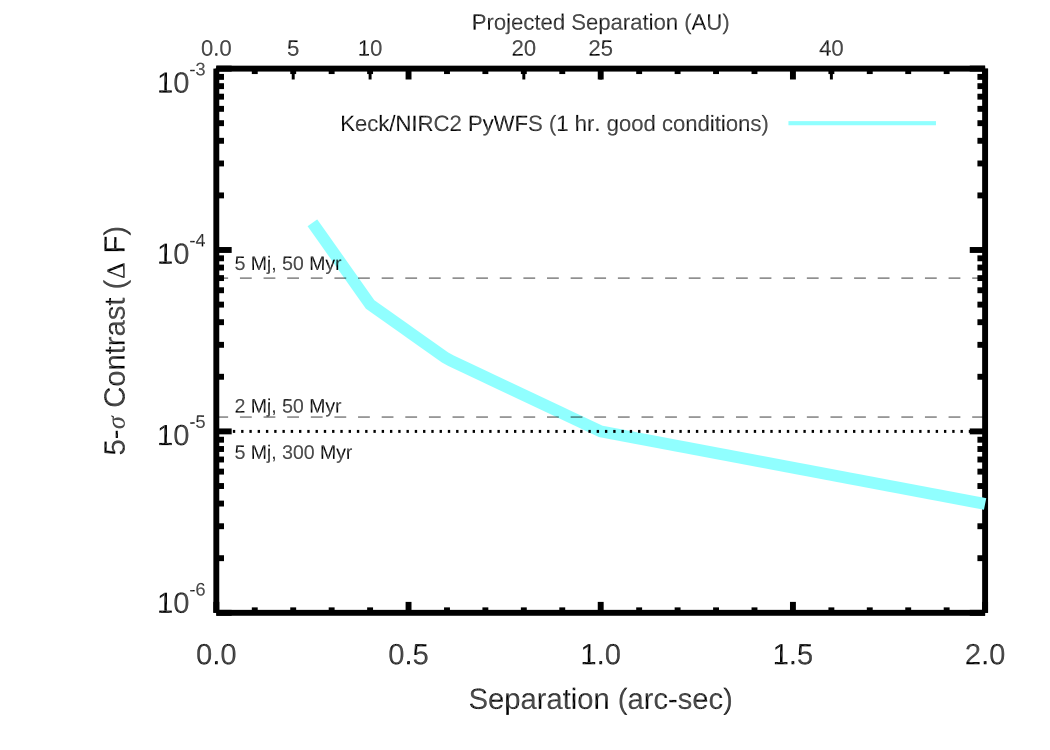}
   %\includegraphics[scale=0.65,clip]{Figures/hd1160comp.png}
  % \end{tabular}
   \end{center}
   \vspace{-0.1in}
   \caption
%>>>> use \label inside caption to get Fig. number with \ref{}
   { \label{fig:contrast} 
   (left) SCExAO/CHARIS 5-$\sigma$ measured (magenta) and soon-to-be expected (green) contrast curves scaled to one hour integration time (magenta) under good conditions with excellent ($\Delta$PA $>$ 50$^{o}$) parallactic angle motion.  
   %after the upgrade of AO-188 to a higher-order DM with an EMCCD used for wavefront sensing and better correction of non-common path aberrations.    
   Horizontal dashed lines note the contrasts expected for 2--5 $M_{\rm J}$ jovian planets around a 50 Myr-old Sun-like star at a distance of 25 $pc$\cite{Baraffe2003}; the dotted line shows contrasts expected for a 5 $M_{\rm J}$, 300 Myr old planet.
   }
   
   %Selected reduction steps in the CHARIS Data Post-Processing Pipeline.  (top-left) A CHARIS data cube aftersky-subtraction and precise image registration have been performed.  (top-right) Kurucz model atmosphere for an F8V star (appropriate for HD 33632 Ab) binned to CHARIS's spectral resolution and resampled along the CHARIS broadband mode wavelength grid.    (bottom-left) ADI/A-LOCI + classical SDI subtracted image of HR 8799 showing extremely high SNR detections of HR 8799 cde.  (bottom-right) Grid of synthetic L-type planets forward-modeled through our data using the approach of Ref. \citenum{Currie2018a} to simulate signal loss due to A-LOCI in ADI+SDI mode.}
\end{figure}

\section{SCExAO/CHARIS and Keck/NIRC2 High-Contrast Sensitivity to Planets as a Function of Age}
Figure \ref{fig:contrast} (left panel) compares performance achieved with SCExAO/CHARIS under good conditions in one hour of integration time with contrasts needed to detect 2--5 $M_{\rm J}$ planets at 50 Myr and 500 Myr.   At angular separations of 0\farcs{}25 and 0\farcs{}5-- 1\farcs{}0 SCExAO/CHARIS achieves contrasts of 10$^{-5}$ and 10$^{-6}$, respectively.   The early 2022 replacement of AO-188 with a 3200 actuator DM for a first-order correction coupled with a fast high-order wavefront sensor will eventually result in a factor of 5--10 gain in raw contrast or 10$^{-6}$ at 0\farcs{}25 for a one-hour sequence with SCExAO/CHARIS utilizing post-processing.   For a 50 Myr old Sun-like star at 25 $pc$, SCExAO/CHARIS is sensitive to 2 $M_{\rm J}$ planets at Saturn-or-greater projected separations and 5 $M_{\rm J}$ planets at a Jupiter-like projected separation.   By mid 2022, SCExAO/CHARIS would be able to detect $\lesssim$2 $M_{\rm J}$ planets at separations approaching that of Jupiter and lower-mass planets at Saturn to Neptune-like separations. 

Jovian planets cool and contract with age: the peak of their thermal emission shifts further into the mid IR at older ages (e.g. \citenum{Skemer2014}).   For intermediate-aged stars -- 300 Myr to 1 Gyr -- jovian exoplanet detectability becomes more favorable with thermal IR platforms like Keck/NIRC2 than near-IR focused ones like SCExAO/CHARIS, even if the former's AO performance is poorer.   Thus, our survey includes Keck/NIRC2 $L_{\rm p}$ coronagraphic imaging obtained with the near-IR PyWFS to complement SCExAO/CHARIS observations (Figure \ref{fig:contrast}, right panel).   

At small angular separations ($\rho$ $\lesssim$ 0\farcs{}5), where planet detection is contrast limited, SCExAO/CHARIS outperforms NIRC2 $L_{\rm p}$ imaging even for stars up to 300--500 Myr old.   However, at wider separations or for older ages, NIRC2 data imaging can detect lower-mass planets.   NIRC2 also allows the identification of substellar candidates outside the CHARIS field of view, which can still correspond to solar system-like scales for the nearest stars.   The near-IR PyWFS installed at Keck also allows NIRC2 observations to effectively probe optically fainter, later-type stars than typically possible with SCExAO/CHARIS right now.

\section{Target Selection} 
Our target selection focuses on stars showing statistically significant accelerations whose distances are small enough and ages young enough to suggest the systems may have imageable substellar companions at angular separations accessible by CHARIS and NIRC2.

\subsection{Acceleration and Mass}
The HGCA computes accelerations from deviations between three proper motion measurements: \textit{Hipparcos} proper motions (epoch $\sim$ 1991.25), \textit{Gaia} proper motions (epoch $\sim$ 2015.5), and the \textit{Gaia--Hipparcos} positional difference divided by the $\sim$ 25 year time baseline between these two measurements (the ``scaled positional difference")\cite{Brandt2019}.   The Hipparcos proper motions draw from a linear combination of two separate reductions\cite{ESA1997,vanLeeuwen2007}: HGCA applies other cross-calibrations (e.g. rotation, local offsets between reference frames, etc.).   The most precise astrometric measurement is the \textit{Gaia--Hipparcos} positional difference.   The acceleration, $a$, at weighted epoch $t_{\rm a}$ is then given as\cite{Brandt2019}:
\begin{equation}
a_{\alpha, \delta}[t_{\rm a}] = \frac{\Delta\mu_{Gaia}}{(t_{Gaia}-t_{Hip})/2}, \, \, \rm{where}
\end{equation}
\begin{equation}
t_{\rm a} = \frac{3t_{Gaia}+t_{Hip}}{4}.
\end{equation}
The $\chi^{2}$ statistic to quantify the statistical significance of the acceleration considers the full covariance between the \textit{Hipparcos} and \textit{Gaia} measurements: i.e. 
\begin{equation}
\chi^{2} = [\Delta~\mu_{\alpha\star},\Delta~\mu_{\delta}]\mathbf{C}^{-1}\begin{bmatrix} \Delta~\mu_{\alpha\star}\\\Delta~\mu_{\delta}, \end{bmatrix}
\end{equation}
where the covariance matrix \textbf{C} is modified to consider error inflation in the \textit{Hipparcos} data and a systematic proper motion uncertainty.   For two degrees of freedom, a 2, 3, and 5-$\sigma$ acceleration corresponds to $\chi^{2}$ = 6.2, 11.8, and 28.75.  

For an angle $\phi$ between the position vector between the star and companion, $b$, and the sky plane and an absolute separation of $r$, the astrometric acceleration is related to the companion mass by:
\begin{equation}
a_{\alpha\delta} = \frac{GM_{b}}{r^{2}}cos\phi, 
\end{equation}
where $r$ is related to $\phi$, the parallax ($\bar{w}$), and angular separation ($\rho$) by $\rho$ = $r\bar{w}cos\phi$. 

Absolute astrometry from HGCA gives precise estimates for $a_{\alpha\delta}$; the vector for the proper motion anomaly from HGCA coupled with relative astrometry of the companion place limits on $\phi$ and thus $M_{\rm b}$.    While the addition of radial-velocity (RV) data place even stronger companion mass limits\cite{Brandt2019}, in many cases RV is not necessary to constrain the companion mass to $\sim$ 10--20\%, especially given multi-epoch relative astrometry from direct imaging (see \S{6}, \citenum{Currie2020a}).

To screen for accelerating stars with candidate substellar companions, we used the acceleration defined from comparing the proper motion from \textit{Gaia} to the \textit{Gaia}--\textit{Hipparcos} positional difference divided by the time baseline.   We estimate a lower limit on the mass of an accelerating star, assuming an orbit viewed perfectly face on and at angular separations of $\rho$ = 0\farcs{}25--0\farcs{}75.   Systems with $M_{\rm b}$ $>$ 200 $M_{\rm J}$ at 0\farcs{}5 and those without at least a marginally-significant acceleration ($>$2-$\sigma$) are removed from the sample.
%An initial screening of a sample of targets showed that those with extremely high $\chi^{2}$ values ($\chi^{2}$ $>$ 1000)
%(hereinafter referredto as the scaled positional difference).
%statistically significant.
%mass equation

\subsection{Age and Distance}
\begin{figure}[ht]
   \begin{center}
  % \begin{tabular}{c} %% tabular useful for creating an array of images 
  \centering
   \includegraphics[scale=0.7,trim = 0mm 0mm 10mm 0mm,clip]{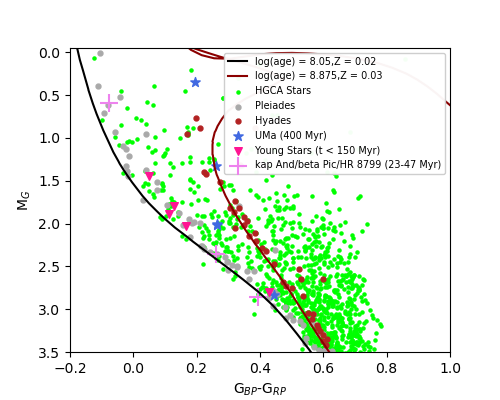}
   \includegraphics[scale=0.575,trim = 0mm 0mm 15mm 0mm,clip]{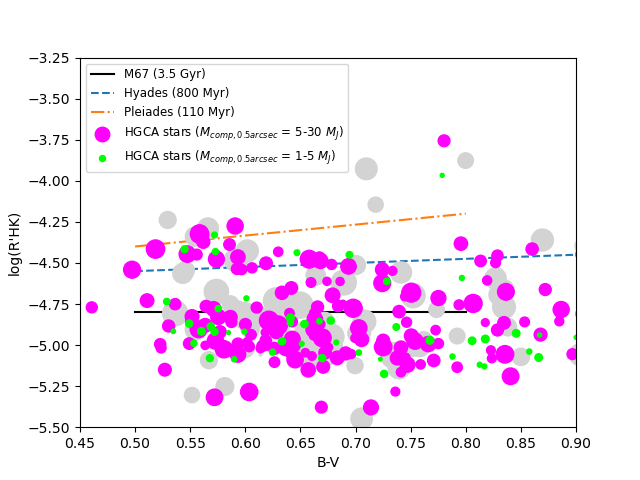}
   %\includegraphics[scale=0.65,clip]{Figures/hd1160comp.png}
  % \end{tabular}
   \end{center}
   \vspace{-0.1in}
   \caption
%>>>> use \label inside caption to get Fig. number with \ref{}
   { \label{fig:ageselect} 
   (left) HR diagram using \textit{Gaia} data comparing the positions for our BAF sample members and Pleiades, Hyades, interferometrically-observed UMa stars, other young stars, and isochrones appropriate for the Pleiades and Hyades.  Stars lying to the left of the Hyades are retained for observations; those lying on/near the Pleiades locus are prioritized. (right) Activity vs B-V color for for a subset of solar-type sample members compared to median activity levels in the Hyades, Pleiades, and M67.   Symbol sizes are proportional to the minimum companion mass if it were located at 0\farcs{}5 from the host star: grey symbols denote likely binary companions or very close-in substellar companions ($\rho$ $\lesssim$ 0\farcs{}25).}
   %The posterior distribution of masses for HD 33632 Ab using \textit{Gaia} DR2 astrometry as in \citenum{Currie2020a} vs. \textit{Gaia}-eDR3 astrometry: the updated HGCA catalog uses \textit{Gaia}-eDR3 astrometry, yielding a $\sim$ 50\% smaller mass uncertainty (G. M. Brandt 2021, submitted).   (right) HD 33632 Ab's mass posterior distribution derived from a single epoch of high-contrast imaging data (2018; grey) and the two epochs reported in \citenum{Currie2020a}.   }
\end{figure}
RV surveys suggest that the frequency of jovian planets peaks at $\sim$ 3 au, near the water ice line for Sun-like stars, and drops significantly by 30 au \cite{Fulton2021}.   Our detection capabilities sensitively depend on stellar age (see Figure \ref{fig:contrast}), favoring stars younger than 300 Myr.  We use this information to focus on stars whose unseen companions could be imageable.

For B, A, and early F spectral types, the stars' Hertzsprung-Russell (HR) diagram positions drawn from \textit{Gaia} photometry help select for youth.   Figure \ref{fig:ageselect} compares the HR diagram positions for BAF stars in the Pleiades ($t$ $\sim$ 115 Myr) and Hyades (700-800 Myr), PARSEC isochrones that reproduce the Pleiades and Hyades loci\cite{Bressan2012,Gaia2018} and young stars with ages derived from optical interferometry, including those in the 400 Myr-old Ursa Majoris (UMa) moving group \cite{Jones2015,Jones2016}.   Well-known planet hosts $\beta$ Pic and HR 8799 ($\sim$ 23 Myr and 40 Myr) plot below the Pleiades locus, while $\kappa$ And plots above the locus despite being less than half the age of the Pleiades (47 Myr).  Stars in UMa (blue stars) are just slightly blueward of the Hyades locus.  

Many stars showing an acceleration plausibly due to an unseen companion plot below/blueward of the Ursa Majoris and Hyades loci, suggesting that they are younger than 400 Myr.    Stars with HR diagram positions between UMa and the Pleiades could have ages between 100 Myr and 400 Myr or possibly younger than 100 Myr due to rotation/oblateness.   The plot shows a substantial number of sample members on the Pleiades sequence, strongly indicative of a population of young, accelerating stars.

Young solar-type stars (e.g. B-V = 0.5--0.9) show significantly elevated Ca$_{\rm HK}$ emission compared to the Sun; the equivalent width of these lines declines as a function of age \cite{Mamajek2008}.   To identify solar-type stars younger than the Hyades, we cross-referenced the list of accelerating with various catalogues reporting log(R$_{\rm HK}$)(e.g. \citenum{Pace2013}) and the Tycho II catalog reporting $B_{\rm T}$ and $V_{\rm T}$ and converted photometry to standard Johnson-Cousins photometric system (Figure \ref{fig:ageselect}, right panel).  For later, redder stars we utilized lithium equivalent width measurements (e.g. \citenum{Guillout2009}).

\subsection{Previous Data}
Finally, we use previously acquired data to screen our remaining targets for stars with known moderate to equal mass stellar companions.   Primary data sources include the Keck Observatory Archive (KOA), previous direct imaging surveys\cite{Galicher2016}, and \textit{Simbad} database information.   Common interloping objects include wide separation equal mass systems (for FGK stars) and spectroscopic binaries (for early-type stars).   For some targets, we reduced unpublished high-contrast imaging data from the KOA and other sources to verify that they lacked a stellar companion using a well-tested general purpose pipeline\cite{Currie2011}.

\begin{figure}[ht]
   \begin{center}
  % \begin{tabular}{c} %% tabular useful for creating an array of images 
 % \centering
   \includegraphics[scale=0.375,clip]{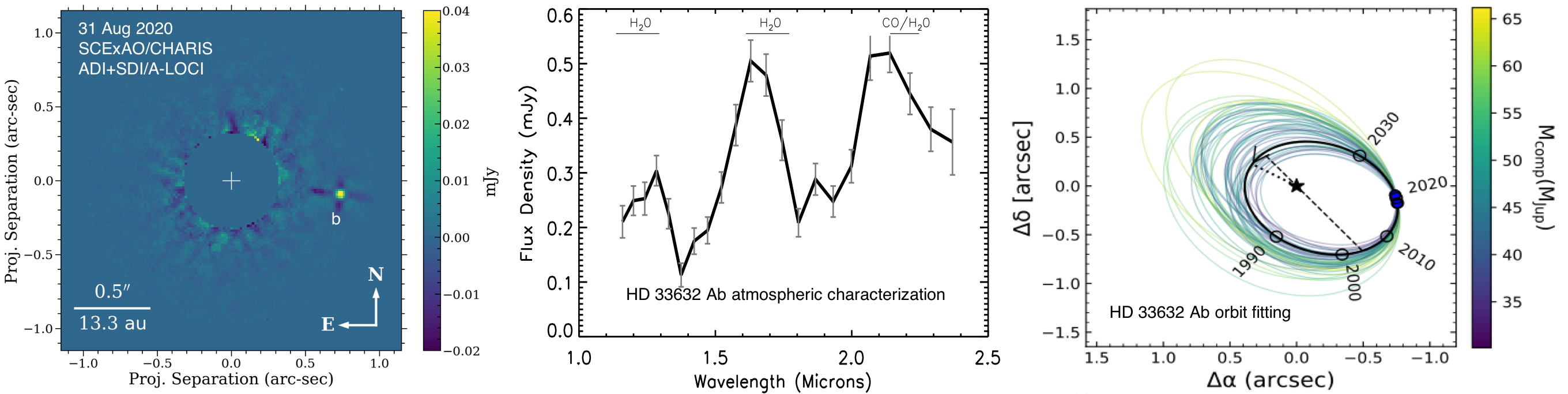}
  % \includegraphics[scale=0.45,clip]{Figures/HD_33632_with_and_without_2020relast_option2.png}
   %\includegraphics[scale=0.65,clip]{Figures/hd1160comp.png}
  % \end{tabular}
   \end{center}
   \vspace{-0.1in}
   \caption
%>>>> use \label inside caption to get Fig. number with \ref{}
   { \label{fig:hd33632} 
   Detection and characterization of HD 33632 Ab from Currie et al. \citenum{Currie2020a}.   The left panel shows recovery of HD 33632 Ab from 31 August 2020 data processed with a combination of ADI and SDI, the middle panel shows the spectrum of HD 33632 Ab with wavelength ranges for major molecular species shown, and the right panel shows a simultaneous fit to the orbit and dynamical mass for HD 33632 Ab using \textit{Gaia-DR2} astrometry.     }
\end{figure}

\begin{figure}[ht]
   \begin{center}
  % \begin{tabular}{c} %% tabular useful for creating an array of images 
  \centering
   \includegraphics[scale=0.45,clip]{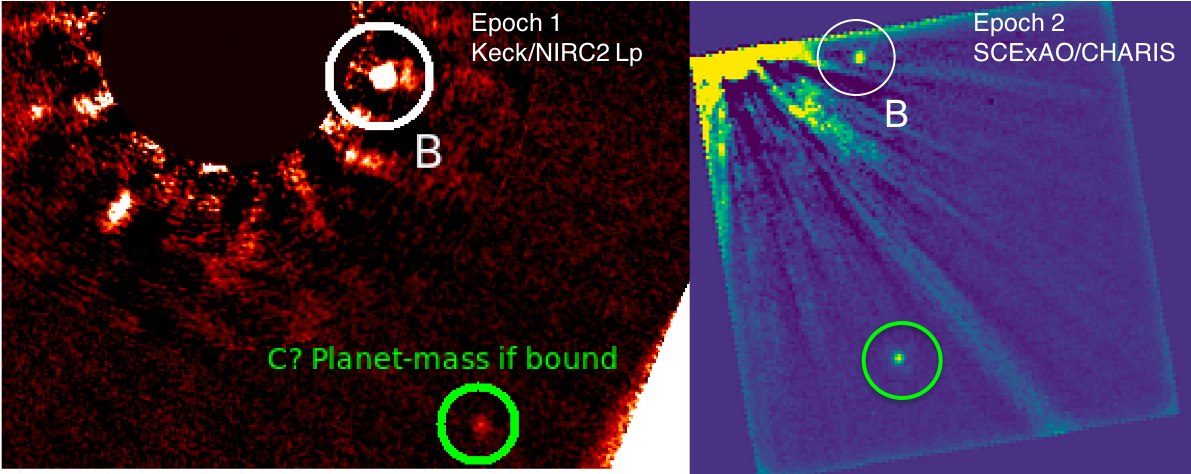}
  % \includegraphics[scale=0.45,clip]{Figures/HD_33632_with_and_without_2020relast_option2.png}
   %\includegraphics[scale=0.65,clip]{Figures/hd1160comp.png}
  % \end{tabular}
   \end{center}
   \vspace{-0.1in}
   \caption
%>>>> use \label inside caption to get Fig. number with \ref{}
   { \label{fig:hd33632c} 
   NIRC2 (November 2018) and offsetted CHARIS (August 2020) data at a more aggressive image stretch showing the detection of a fainter, wider separation point source.   Subsequent analysis suggests that this object is likely a background star despite being detected in $L_{\rm p}$.    }
\end{figure}
\section{Discoveries from the First Full Year of the HGCA Survey}
\textbf{HD 33632 Ab}\cite{Currie2020a} -- The first discovery from our SCExAO/CHARIS HGCA survey is HD 33632 Ab, a substellar companion to a nearby mature (1--2.5 Gyr old) Sun-like star, HD 33632 Aa.  We image the companion in two epochs -- October/November 2018 and August/September 2020 -- imaged at a projected separation of $\sim$ 20 au (Figure \ref{fig:hd33632}).   HD 33632 Ab's astrometry showed clear evidence for counterclockwise orbit motion.  

We used HD 33632 Ab's CHARIS $JHK$ spectrum and NIRC2 $L_{\rm p}$ photometry to constrain its atmospheric properties.   HD 33632 Ab matches field brown dwarfs lying at the L/T dwarf transition tracing the dissipation of clouds below substellar atmospheres.  Combining absolute astrometry of the star from \textit{Hipparcos} and \textit{Gaia} with relative astrometry from CHARIS and NIRC2 and some RV data constrained HD 33632 Ab's dynamical mass.   Using the Data Release 2 astrometric measurements for \textit{Gaia}, we compute an inferred mass of $\sim$ 46 $M_{\rm J}$ $\pm$ 8 $M_{\rm J}$.   While this is clearly about the deuterium-burning limit that often is invoked to separate planets from brown dwarfs, best-fit orbital parameters for HD 33632 Ab favor extremely low eccentricities that are more characteristic of imaged exoplanets\cite{Bowler2020}.

Additionally, we identified a second candidate companion around HD 33632 Ab in our NIRC2 data at a projected separation of $\rho$ $\sim$ 2\farcs{}25 (Figure \ref{fig:hd33632c}).   If a bound companion, this object's $L_{\rm p}$ brightness would be consistent with a superjovian companion at $\sim$60 au.   However, follow-up data with SCExAO/CHARIS showed a spectrum inconsistent with a cool substellar companion and astrometry consistent with a background star.   
%This analysis shows the utility of multi-wavelength, multi-epoch 

\textbf{HIP 109427 and HD 91312} -- In Steiger et al. \citenum{Steiger2021}, we reported the discovery of HIP 109427 B, a low-mass stellar companion inducing an acceleration on its A type host star.  The discovery of HIP 109427 B drew from SCExAO/CHARIS high-contrast spectroscopy, NIRC2 $L_{\rm p}$ imaging, and SCExAO/MEC \cite{Walter2020} using stochastic speckle discrimination (SSD).  Joint fitting of HIP 109427 B's relative astrometry and the host star's absolute astrometry constrained the companion to have a mass of M $\sim$ 0.280$^{+0.180}_{-0.059}$ $M_{\rm \odot}$.  The discovery demonstrated both the HGCA survey concept as well as the efficacy of SSD to identify low-mass companions near the telescope diffraction limit.

In Chilcote et al. (2021, under review), we report the direct imaging discovery of a low-mass stellar companion to the nearby A7 star HD 91312.   The companion, HD 91312 B, was inferred from a long-term radial-velocity trend implying the existence of a companion beyond $\sim$ 1 au from the star.   HD 91312 B has a nearly edge on orbit with a Saturn-like semimajor axis.   From combining direct imaging, astrometry, and radial-velocity data, we constrain the companion mass to within $\sim$12\%. 

\begin{figure}[ht]
   \begin{center}
  % \begin{tabular}{c} %% tabular useful for creating an array of images 
  \centering
   \includegraphics[scale=0.75,clip]{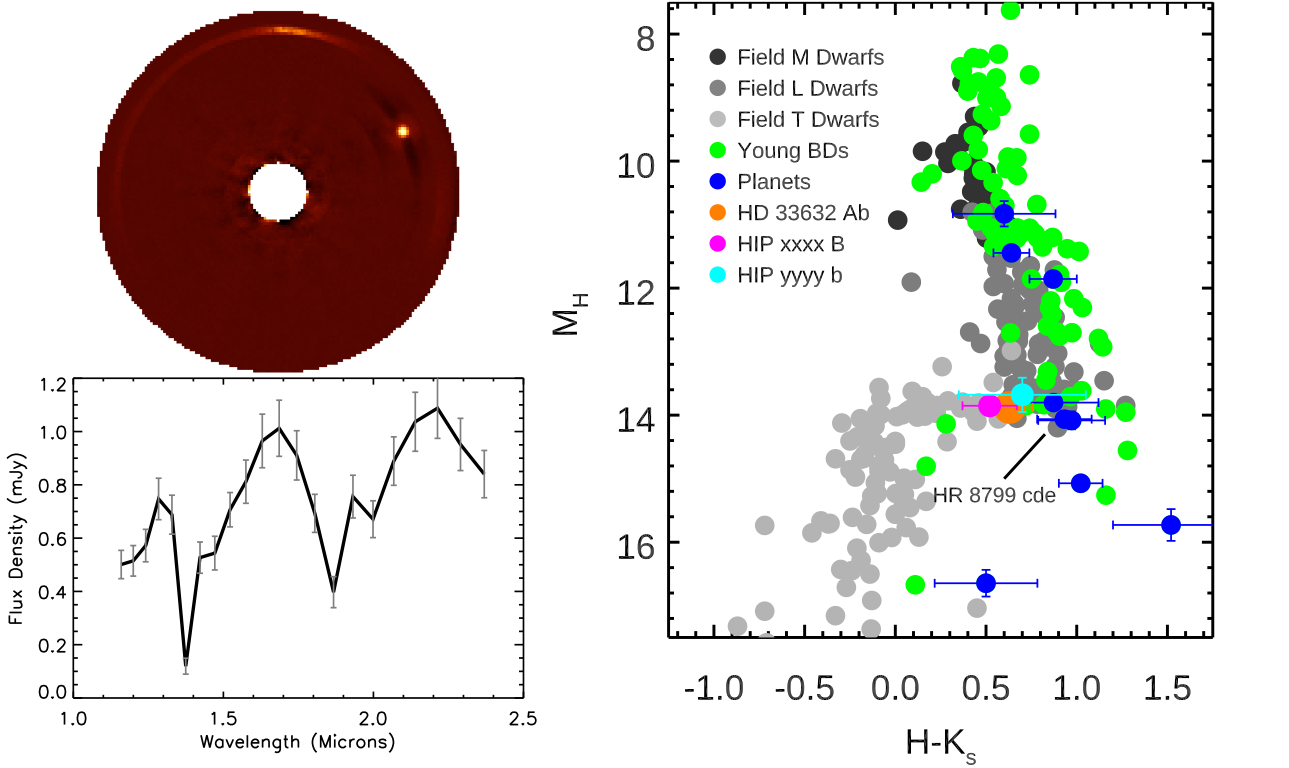}
  % \includegraphics[scale=0.45,clip]{Figures/HD_33632_with_and_without_2020relast_option2.png}
   %\includegraphics[scale=0.65,clip]{Figures/hd1160comp.png}
  % \end{tabular}
   \end{center}
   \vspace{-0.1in}
   \caption
%>>>> use \label inside caption to get Fig. number with \ref{}
   { \label{fig:newcand} 
   (left) Discovery image (top) and spectrum (bottom) for a new candidate substellar companion identified from this survey.  (right) $H$/$H$-$K_{\rm s}$ color-magnitude diagram showing the positions of three companions discovered from our survey compared to the field L/T dwarf sequence\cite{DupuyLiu2012}, young substellar objects, and directly imaged planets including HR 8799 bcde.   }
\end{figure}

\textbf{New Substellar Companions/Candidates and a Candidate Directly-Imaged Exoplanet} --  Our survey has identified over 10 candidate/likely low-mass companions around nearby stars, a subset of which are or could be substellar.  Figure \ref{fig:newcand} (left, middle panels) shows one candidate brown dwarf around an early type star identified from now imaged with both SCExAO/CHARIS and NIRC2 (T. Uyama et al. in prep.).   Its spectrum best resembles objects near the M/L dwarf transition and requires a second epoch detection to confirm common proper motion and better constrain the companion mass.   Another star shows a faint brown dwarf companion with a spectrum extremely similar to that of HD 33632 Ab, although the companion is plausibly lower in mass (M. Kuzuhara et al. in prep.).

Finally, our survey has identified a very low-mass substellar companion and candidate planet around a nearby, young dusty star.   The companion's common proper motion is confirmed; analysis of recently acquired astrometry will better constrain its dynamical mass.   
Figure \ref{fig:newcand} (right panel) shows an $H$/$H$-$K_{\rm s}$ color magnitude diagram showing the positions of three objects discovered from our survey: HD 33632 Ab, one of our brown dwarf candidates, and the candidate planet.   The three companions have different ages and masses but extremely similar near-IR colors, all lying near the L/T transition and close in color-magnitude space to HR 8799 cde.   A joint analysis of these objects will better clarify the atmospheric evolution of substellar companions near the L/T dwarf transition as a function of mass and age.

\section{Dynamical Masses}
Direct imaging surveys using a blind target selection rely on luminosity evolution models to convert from planet brightness to mass.   In addition to uncertainties in the luminosity evolution itself \cite{Spiegel2012}, this approach requires precise stellar ages, which are often unavailable, especially if the star is not a member of a young moving group or open cluster.   A survey utilizing RV or astrometry for target selection is fundamentally different, as it allows \textit{direct} dynamical mass constraints.   
The HGCA survey constrains  planet dynamical masses directly by jointly modeling CHARIS and NIRC2 relative astrometry with absolute astrometry of the star (and sometimes RV data).  

\begin{figure}[ht]
   \begin{center}
  % \begin{tabular}{c} %% tabular useful for creating an array of images 
  \centering
   \includegraphics[scale=0.45,clip]{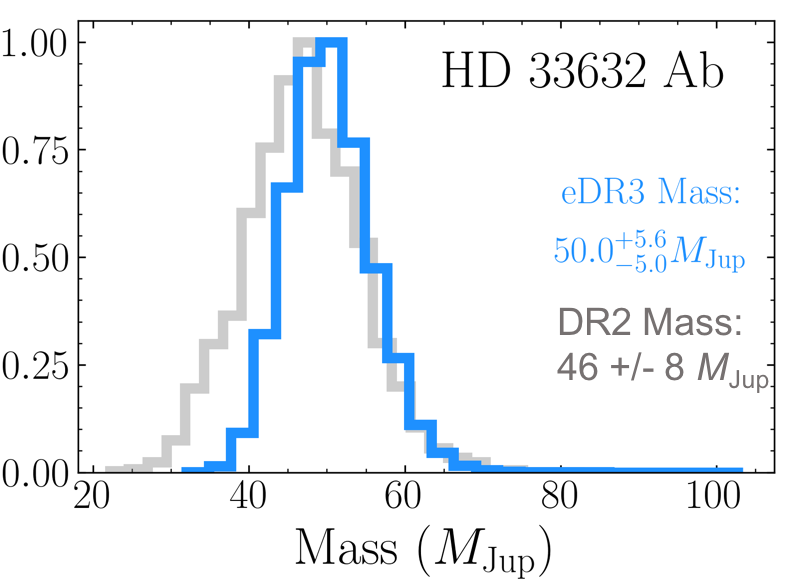}
   \includegraphics[scale=0.45,clip]{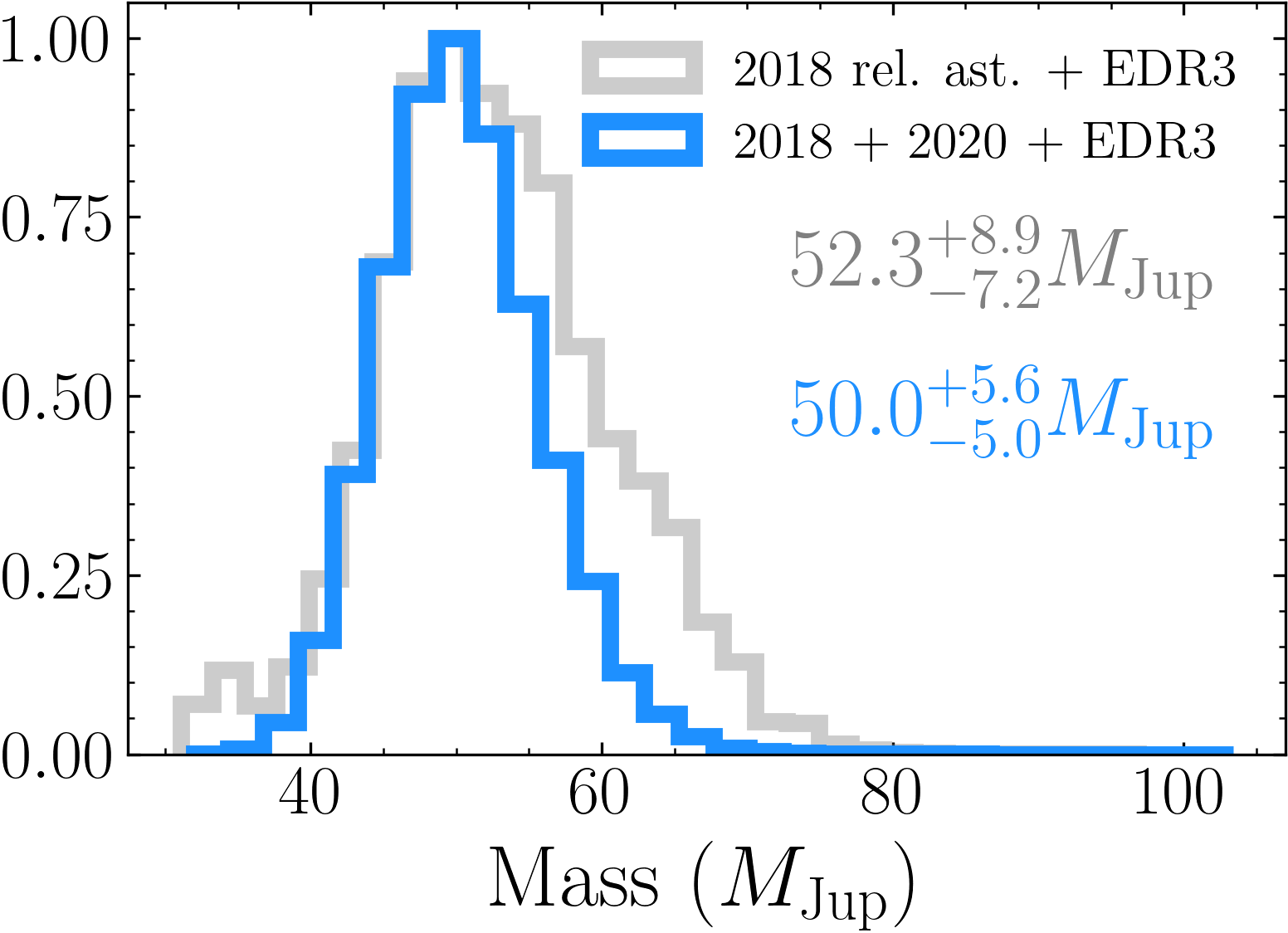}
   %\includegraphics[scale=0.65,clip]{Figures/hd1160comp.png}
  % \end{tabular}
   \end{center}
   \vspace{-0.1in}
   \caption
%>>>> use \label inside caption to get Fig. number with \ref{}
   { \label{fig:massconstraints} 
   (left) The posterior distribution of masses for HD 33632 Ab using \textit{Gaia} DR2 astrometry as in \citenum{Currie2020a} vs. \textit{Gaia}-eDR3 astrometry: the updated HGCA catalog uses \textit{Gaia}-eDR3 astrometry, yielding a $\sim$ 50\% smaller mass uncertainty (G. M. Brandt 2021, submitted).   (right) HD 33632 Ab's mass posterior distribution derived from a single epoch of high-contrast imaging data (2018; grey) and the two epochs reported in Currie et al. \citenum{Currie2020a}.   }
\end{figure}

Figure \ref{fig:massconstraints} explores dynamical mass constraints on companions imaged from our survey.   We started the HGCA survey using DR2 for the \textit{Gaia} absolute astrometry source: these data helped constrain the mass of HD 33632 Ab in Currie et al. \citenum{Currie2020a}.  As described in Brandt et al. \citenum{Brandt2021}, the \textit{Gaia} Early Data Release 3 (eDR3) provides a substantial improvement in astrometric precision.   For HD 33632 Ab, eDR3 astrometry shrink the astrometric errors for HD 33632 Ab by about 50\% (G. M. Brandt et al. 2021, submitted).   Because of the long time baseline between \textit{Hipparcos} and \textit{Gaia} and the extremely precise astrometry from \textit{Gaia}, even a single relative astrometric point from SCExAO/CHARIS helps to provide useful constraints on a companion's dynamical mass.  For HD 33632 Ab, a single CHARIS astrometric point, when coupled with absolute astrometry from eDR3 (right panel), yields precision similar to that achieved with DR2 from two epochs of CHARIS relative astrometry (left panel).

\section{Future Directions}
Approved and recently funded upgrades to SCExAO will substantially improve the planet detection and characterization potential from this survey.
Previous SPIE submissions describe the full range of upgrades planned for SCExAO in the next several years \cite{Currie2020b,Guyon2020a}.     Key technical improvements to SCExAO particularly relevant to our HGCA survey focus on wavefront sensing and control --  the replacement of AO-188 with a 3200-actuator deformable mirror, near-IR PyWFS, and high-order visible WFS -- will provide extreme AO capability with the first AO correction stage: the current SCExAO WFC loop can further sharpen this correction to achieve exceptionally high Strehl ratios and/or perform focal-plane wavefront sensing and control methods to remove quasi-static speckles near the optical axis.   

\begin{figure}[h]
\vspace{-0.1in}
   \begin{center}
  % \begin{tabular}{c} %% tabular useful for creating an array of images 
  \centering
   \includegraphics[scale=0.67,clip]{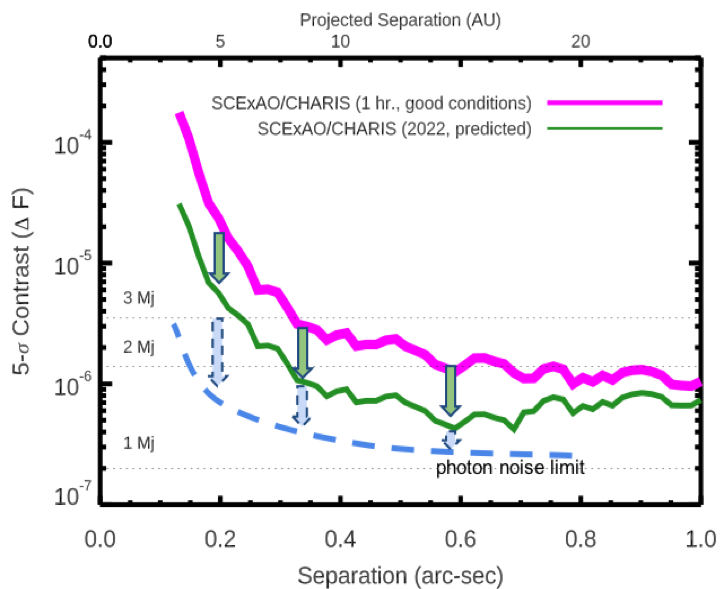}
  %   \includegraphics[scale=0.47,clip]{Figures/nirc2_one_hour_contrast.eps}
   %\includegraphics[scale=0.65,clip]{Figures/hd1160comp.png}
  % \end{tabular}
   \end{center}
   \vspace{-0.1in}
   \caption
%>>>> use \label inside caption to get Fig. number with \ref{}
   { \label{fig:eventualcontrast} 
   SCExAO/CHARIS 5-$\sigma$ measured (magenta) and soon-to-be expected (green) contrast curves scaled to one hour integration time under good conditions compared to the photon noise limit (blue dashed line).   Improved calibration of the speckle halo will push system performance closer to the photon noise floor, allowing the detection of young versions of Jupiter.  
   %with excellent ($\Delta$PA $>$ 50$^{o}$) parallactic angle motion.  
   %after the upgrade of AO-188 to a higher-order DM with an EMCCD used for wavefront sensing and better correction of non-common path aberrations.    
   %Horizontal dashed lines note the contrasts expected for 2--5 $M_{\rm J}$ jovian planets around a 50 Myr-old Sun-like star at a distance of 25 $pc$\cite{Baraffe2003}; the dotted line shows contrasts expected for a 5 $M_{\rm J}$, 300 Myr old planet.
   }
   
   %Selected reduction steps in the CHARIS Data Post-Processing Pipeline.  (top-left) A CHARIS data cube aftersky-subtraction and precise image registration have been performed.  (top-right) Kurucz model atmosphere for an F8V star (appropriate for HD 33632 Ab) binned to CHARIS's spectral resolution and resampled along the CHARIS broadband mode wavelength grid.    (bottom-left) ADI/A-LOCI + classical SDI subtracted image of HR 8799 showing extremely high SNR detections of HR 8799 cde.  (bottom-right) Grid of synthetic L-type planets forward-modeled through our data using the approach of Ref. \citenum{Currie2018a} to simulate signal loss due to A-LOCI in ADI+SDI mode.}
\end{figure}

In addition to the AO-188 upgrade, SCExAO will undergo key improvements in its wavefront sensing and control capabilities that will push its performance closer to the photon noise floor (Figure \ref{fig:eventualcontrast}).  Advances include self-calibrating wavefront sensing and control to identify speckles using WFS telemetry (Currie et al. \citenum{Currie2020b}, O. Guyon, 2021 SPIE proceedings), better compensating for non-common path aberrations using a direct reinforcement wavefront heuristic optimization (DrWHO) (N. Skaf, 2021 SPIE proceedings), wavefront sensing using non-redundant aperture masking interferometry (V. Deo, 2021 SPIE proceedings), and photonics technology for wavefront sensing/control and science capabilities (S. Vievard, 2021 SPIE proceedings).

Subsequent \textit{Gaia} data releases will improve the mission's achievable astrometric precision to the $\mu$-arcsecond level and more decisively identify companions around nearby stars that may be substellar and imageable.   Multiple epochs of \textit{Gaia} data will provide far better constraints on companions already found or soon-to-be identified through eDR3.   

Finally, our survey prefigures the kind of exoplanet direct imaging and characterization program that will be carried out with future 30m-class telescopes on the ground and NASA missions.   As shown in Brandt et al. \citenum{Brandt2019astro}, a single astrometric measurement from the Roman-WFI instrument coupled with extreme AO imaging from a 30m-class telescope should yield the detection and spectral characterization of over 150 nearby planets spanning a range of temperatures and orbits.  High-contrast imaging instruments such as \textit{METIS} on the \textit{European Extremely Large Telescope} or the \textit{Planetary Systems Imager} on the \textit{Thirty Meter Telescope} would be able to image and spectrally characterize planets identified from Roman-WFI \cite{Brandl2014,Fitzgerald2019}.

\acknowledgments % equivalent to \section*{ACKNOWLEDGMENTS}       
\indent The authors acknowledge the very significant cultural role and reverence that the summit of Mauna Kea holds within the Hawaiian community.  We are most fortunate to have the opportunity to conduct observations from this mountain.   We support and endeavor to contribute to respectful, effective stewardship of cultural, natural, and scientific resources that properly honors these lands.   
%\indent We wish to acknowledge the contributions of the current and recent Subaru and Keck Observatory operator, daycrew, and computer support staff employees, including but not limited to Kianna Schubert, Matthew Wung, Dex Alpiche,  Michael Balberino, Timothy Castro, Jordan Akiona, Lucio Ramos, Michael Lemmon, Rita Morris, Randy Campbell, Heather Hershley, and Andrew Cooper.  Their expertise, ingenuity, and dedication is indispensible to the successful operation of Subaru and Keck: without their contributions, the results presented in this paper would not have been possible.

\indent We acknowledge the critical importance of the current and recent Subaru Telescope daycrew, technicians, support astronomers, telescope operators, computer support, and office staff employees, especially during the challenging times presented by the COVID-19 pandemic.  Their expertise, ingenuity, and dedication is indispensable to the continued successful operation of these observatories.  %Without their contributions, science like that presented in this paper would not have been possible.
\\
\indent We thank the Subaru Time Allocation Committee for their generous support of this program.  TC was supported by a NASA Senior Postdoctoral Fellowship and NASA/Keck grant LK-2663-948181.   TB gratefully
acknowledges support from the Heising-Simons foundation and from NASA under grant \#80NSSC18K0439. 
%MT is supported by JSPS KAKENHI Grant \# 18H05442.
%KW received support provided by NASA through the NASA Hubble Fellowship grant HST-HF2-51472.001-A awarded by the Space Telescope Science Institute, which is operated by the Association of Universities for Research in Astronomy, Incorporated, under NASA contract NAS5-26555.   
%The results reported herein benefited from collaborations and/or information exchange within NASA’s Nexus for Exoplanet System Science (NExSS) research coordination network sponsored by NASA’s Science Mission Directorate. 
\\
\indent The development of SCExAO was supported by JSPS (Grant-in-Aid for Research \#23340051, \#26220704 \& \#23103002), Astrobiology Center of NINS, Japan, the Mt Cuba Foundation, and the director's contingency fund at Subaru Telescope.  CHARIS was developed under the support by the Grant-in-Aid for Scientific Research on Innovative Areas \#2302. 
%Some of the data presented herein were obtained at the W. M. Keck Observatory, which is operated as a scientific partnership among the California Institute of Technology, the University of California and the National Aeronautics and Space Administration. The Observatory was made possible by the generous financial support of the W. M. Keck Foundation.
% References
\bibliography{report} % bibliography data in report.bib
\bibliographystyle{spiebib} % makes bibtex use spiebib.bst

\end{document}